\documentclass[runningheads]{llncs}
\usepackage{color}
\usepackage{comment}
\usepackage{graphicx}
\usepackage{adjustbox}
\usepackage{threeparttable}
\usepackage{diagbox} %backslashbox
\usepackage{acronym}
\usepackage[subrefformat=parens,labelformat=parens]{subfig}
\usepackage{cleveref}
\usepackage{booktabs}
\usepackage{pifont}% http://ctan.org/pkg/pifont
\newif\ifwip
\wiptrue
\ifwip
    \includecomment{comment}
\fi
\specialcomment{comment}{\begingroup\bf\color{red}}{\endgroup}
\ifwip
\else
    \excludecomment{comment}
\fi
\graphicspath{{Figures/}}

\begin{document}
\title{Contribution Title\thanks{Supported by organization x.}}
\title{BOFUSS: The Basic OpenFlow Userspace Software Switch}
\title{The Road to BOFUSS: The Basic OpenFlow User-space Software Switch}

%\subtitle{Submission \#xx, xx pages.}
% Eder, Elisa, Joaquin, Zoltan, Davide, Nicola, Carmelo, Christian
%
\author{
Eder Le\~{a}o Fernandes\inst{1}%\orcidID{0000-1111-2222-3333} 
\and
Elisa Rojas\inst{2}%\orcidID{1111-2222-3333-4444} 
\and
Joaquin Alvarez-Horcajo\inst{2}%\orcidID{2222--3333-4444-5555}
\and
Zolt\`{a}n Lajos Kis\inst{3}
\and
Davide Sanvito\inst{4}
\and
Nicola Bonelli\inst{5}
\and
Carmelo Cascone\inst{6}
\and
Christian Esteve Rothenberg\inst{7}
}

\authorrunning{Fernandes et al.}
\institute{
Queen Mary University of London, UK \\
\email{e.leao@qmul.ac.uk}
\and
University of Alcala, Spain \\
\email{\{elisa.rojas,j.alvarez\}@uah.es}
\and
Ericsson, Hungary\\
\email{zoltan.lajos.kis@ericsson.com}
\and
Politecnico di Milano, Italy\\
\email{davide.sanvito@polimi.it}
\and
University of Pisa, Italy\\
\email{nicola@pfq.io}
\and
Open Networking Foundation\\
\email{carmelo@opennetworking.org}
\and
INTRIG, University of Campinas (UNICAMP), Brazil\\
\email{chesteve@dca.fee.unicamp.br}
%\url{http://www.springer.com/gp/computer-science/lncs} 
}
\maketitle              % typeset the header of the contribution
%
%Please provide an abstract of 150 to 250 words. The abstract should not contain any undefined abbreviations or unspecified references.
\begin{abstract}
Software switches are pivotal in the Software-Defined Networking (SDN) paradigm, particularly in the early phases of development, deployment and testing. Currently, the most popular one is Open vSwitch (OVS), leveraged in many production-based environments. However, due to its kernel-based nature, OVS is typically complex to modify when additional features or adaptation is required. To this regard, a simpler user-space is key to perform these modifications. 

In this article, we present a rich overview of BOFUSS, the basic OpenFlow user-space software switch. BOFUSS has been widely used in the research community for diverse reasons, but it lacked a proper reference document. For this purpose, we describe the switch, its history, architecture, uses cases and evaluation, together with a survey of works that leverage this switch. The main goal is to provide a comprehensive overview of the switch and its characteristics. Although the original BOFUSS is not expected to surpass the high performance of OVS, it is a useful complementary artifact that provides some OpenFlow features missing in OVS and it can be easily modified for extended functionality. Moreover, enhancements provided by the BEBA project brought the performance from BOFUSS close to OVS. In any case, this paper sheds light to researchers looking for the trade-offs between performance and customization of BOFUSS.

%Please provide 4 to 6 keywords which can be used for indexing purposes.
\end{abstract}
%TC:break _main_

%
%
%

\acrodef{s}[BOFUSS]{Basic OpenFlow User Space Switch}
\acrodef{api}[API]{Application Programming Interface} 
\acrodef{ewbi}[EWBI]{East/Westbound Interface}
\acrodef{gui}[GUI]{Graphical User Interface}
\acrodef{ibn}[IBN]{Intent-Based Networking}
\acrodef{icn}[ICN]{Information-Centric Networking}
\acrodef{ide}[IDE]{Integrated Development Environment}
\acrodef{lldp}[LLDP]{Link Layer Discovery Protocol}
\acrodef{nbi}[NBI]{Northbound Interface}
\acrodef{odl}[ODL]{OpenDaylight}
\acrodef{of}[OF]{OpenFlow}
\acrodef{ofdp}[OFDP]{OpenFlow Discovery Protocol}
\acrodef{onf}[ONF]{Open Networking Foundation}
\acrodef{onos}[ONOS]{Open Network Operating System}
\acrodef{ovs}[OVS]{Open vSwitch}
\acrodef{poc}[PoC]{Proof-of-Concept}
\acrodef{sbi}[SBI]{Southbound Interface}
\acrodef{sdn}[SDN]{Software-Defined Networking}
\acrodef{ssl}[SSL]{Secure Sockets Layer}
\acrodef{tcp}[TCP]{Transmission Control Protocol}
\acrodef{tlv}[TLV]{Type-Length-Value}
\acrodef{xml}[XML]{eXtensible Markup Language}

\section{Introduction}
\label{intro}

Over the last decade, \ac{sdn} has been enthroned as one of the most groundbreaking paradigms in communication networks by introducing radical transformations on how networks are designed, implemented, and operated~\cite{Kreutz15}. At its foundations, \ac{sdn} data plane devices (aka. switches) are featured with programmable interfaces (e.g., OpenFlow~\cite{of}) exposed to  controller platforms. More specifically, open source software switches are a pivotal piece in the initial phases of research and prototyping founded on  \ac{sdn} principles. 

Due to their wide use, two open source OpenFlow software switches deserve special attention: \ac{ovs}~\cite{Pfaff15} and \ac{s}~\cite{Fernandes14}. Both have different characteristics that make them the best choice for different types of scenarios, research and deployment objectives. \ac{ovs} is probably the most well-known \ac{sdn} switch and used in commercial environments, mostly in SDN-based datacenter networks based on micro-segmentation following an overlay model (cf.~\cite{Kreutz15}).  \ac{s} is commonly seen as a secondary piece of software switch, mostly used for research purposes, \ac{poc} implementations, interoperability tests, among other non-production scenarios. 

In this article, we present the history of  \ac{s}  going through a comprehensive overview of its architecture, applications, and evaluation.  
Let us start the journey by clarifying that \ac{s} is the name we have chosen for this ``late baptism'', since the switch did not have consistently  used official name. Many authors denominate it as \textit{CPqD switch}, being CPqD (Centro de Pesquisa e Desenvolvimento em Telecomunica\c{c}\~oes) the research and development center located in Campinas, Brazil, where it was developed, funded by the Ericsson Innovation Center in Brazil. Hence, the switch has been also referred to as \textit{CPqD/Ericsson switch}, not only for the funding but also for the original code base from an OpenFlow 1.1 version developed by Ericsson Research TrafficLab~\cite{ofsoftswitch11} after forking Stanford OpenFlow 1.0 reference switch/controller implementation~\cite{ofrefsw} developed around 10 years ago. 
\textit{OF13SS} (from OpenFlow 1.3 software switch), or simply \textit{ofsoftswitch13} (following its code name in the GitHub repository~\cite{ofsoftswitch13}), add to the list of names the software artefact is referred to.  
We believe this naming issues can be explained by the lack of an official publication, since the only publication focused on the tool~\cite{Fernandes14}, written in Portuguese, did not introduce a proper name and mainly used the term \textit{OpenFlow version 1.3 software switch}.

Fixing our historical mistake of not having given a proper name (i.e. BOFUSS) to the widely used switch is one of the target contributions of this article. We delve into the switch history and architecture design in Section~\ref{sec:bofuss}. 
Next, Section~\ref{sec:usecases} presents selected use cases, which are later expanded in Section~\ref{sec:research} through an extensive survey of the works (35+) that leverage \ac{s} in their research production. 
We evaluate and benchmark \ac{s} in Section~\ref{sec:evaluation} and, finally, we conclude the article in Section~\ref{sec:future}.

\section{BOFUSS: Basic OpenFlow Userspace Software Switch}
\label{sec:bofuss}

This section first introduces the history and motivation behind the development of \ac{s}, and then presents its design and architecture.

\subsection{Brief History}

Up until the release of the OpenFlow 1.0 standard, there were three OpenFlow switch implementations that provided more or less full compliance with the standard: i) The Stanford Reference OpenFlow Switch~\cite{ofrefsw}, which was developed along with the standardization process and its purpose was to provide a reference to OpenFlow switch behavior under various conditions; ii) The OpenFlow Python Switch (OFPS), which was implemented as part of the OFTest conformance testing framework~\cite{ofps}, meant primarily as a testing framework, and iii) \ac{ovs}~\cite{Pfaff15,ovs}, the most popular and high performance virtual switch with OpenFlow support.

Since the beginning, the OpenFlow standardization process requires that all proposed features are implemented before they are accepted as part of the standard. During the OpenFlow 1.1 standardization work, most of the new feature prototypes were based on \ac{ovs}, mostly on separate branches, independent of each other. 
Unfortunately, standardization only required that each individual new feature worked, instead of looking for a complete and unique implementation of all features, as a continuous evolution of the standard and \ac{sdn} switches. As a result, when OpenFlow 1.1 was published, no implementation was available. While the independent features were implemented, they applied mutually incompatible changes to the core of the \ac{ovs} code, so it was nearly impossible to converge them into a consistent codebase for \ac{ovs} with complete support for OpenFlow 1.1.

This lead to the development of \ac{s}, as already explained in the introduction, popularly known as \textit{CPqD} or {ofsoftswitch13} among other code names. 
The core idea was the need of a simpler implementation to be used for multiple purposes such as: i) a reference implementation to verify standard behavior, ii) an implementation with enough performance for test and prototype deployments, and iii) an elementary base to implement new features with ease.

The code of the switch was based on the framework and tools provided by the Reference OpenFlow Switch. Nevertheless, the datapath was rewritten from scratch to make sure it faithfully represented the concepts of the OpenFlow 1.1 standard. Additionally, the OpenFlow protocol handling was factored into a separate library, which allowed, for example, the implementation of the OpenFlow 1.1 protocol for the NOX controller. The first version of this switch was released in May 2011~\cite{of11ann}.

Afterwards, the software became the first virtual switch to feature a complete implementation of OpenFlow 1.2 and 1.3, showcasing IPv6 support using the OpenFlow Extensible Match (OXM) syntax~\cite{of12toolkitann}. Because of the comprehensive support to OpenFlow features and the simple code base, the switch gradually gained popularity both in academia and in open-source OpenFlow prototyping at the \ac{onf}.

\subsection{Design and Architecture}

The design and implementation of software for virtual switches is typically complex, requiring from developers knowledge of low level networking details. Even though it is hard to escape the intricate nature of software switches, \ac{s} main focus is simplicity. As such, the design and implementation of components and features from the OpenFlow specification seek for easiness to understand and modify. 

The OpenFlow specification does not stipulate data structures and algorithms to implement the pipeline of the switches that support the protocol. As long as the implementation follows the described behavior, there is freedom to define the structure of components. In the design of \ac{s}, whenever possible, the most elementary approach is chosen. Frequently, the straightforward solution is not the most efficient, but exchanging performance for simplicity is a trade-off worth paying, especially when fast prototyping in support of research is prioritized. 

\begin{figure}[!ht]
    \center
    \includegraphics[width=0.8\textwidth, height=10cm]{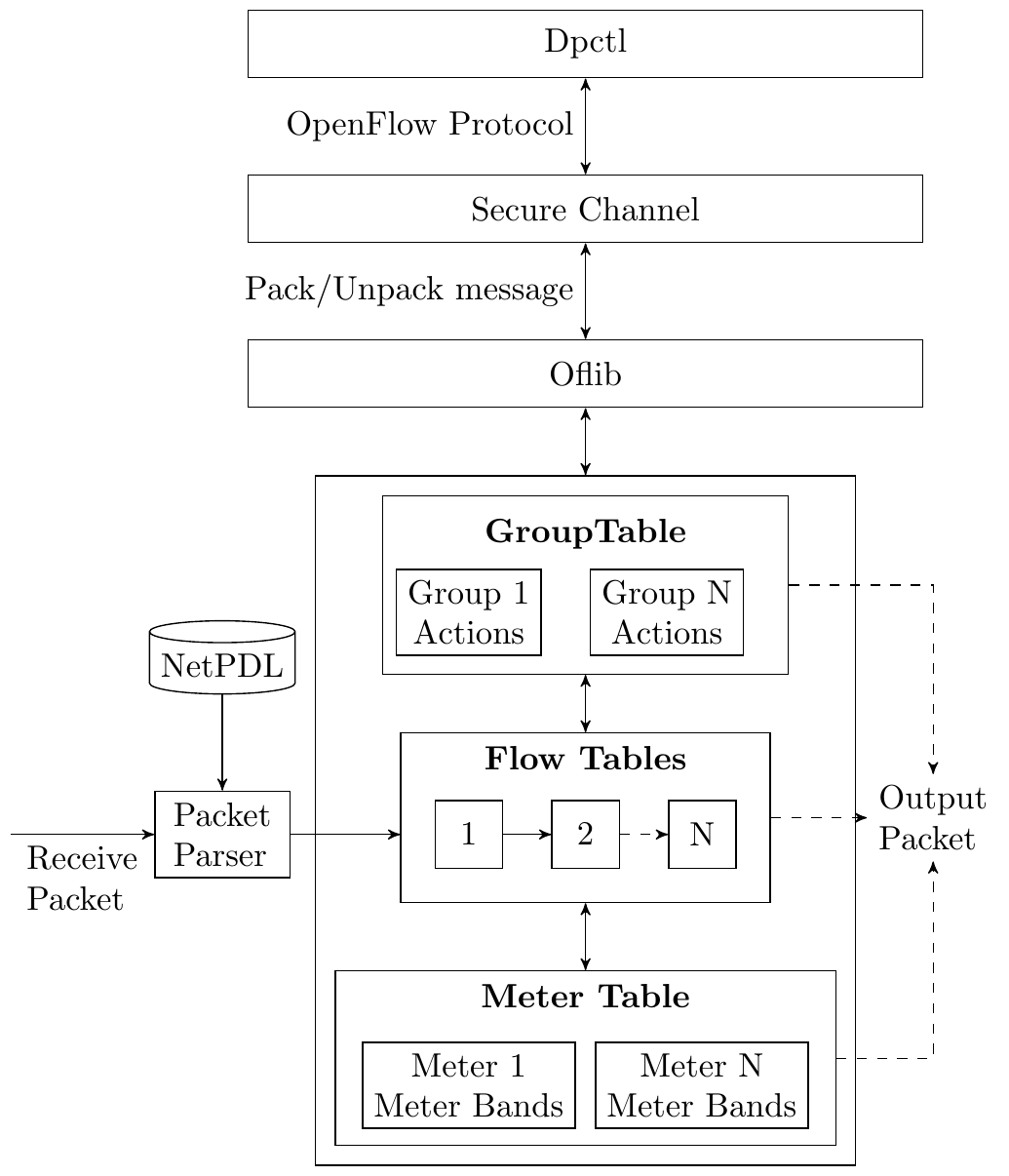}
    \caption{Overview of the architecture from \ac{s}}
    \label{fig:arch}
\end{figure}

We now discuss the structure and organization of \ac{s}, depicted in Figure~\ref{fig:arch}, and how it implements the OpenFlow pipeline. The details presented here aim to be an introduction and starting point for adventuring researchers and developers interested in using \ac{s} to develop and test new features. Appendix~\ref{app:extguide} points to detailed guides that demonstrate how to add or extend switch functionalities.

% Describe architecture and implementation decisions. It would be nice to have 
% pointers to aid people that want to modify the software switch:
% e.g:
% NETPDL simplify packet parsing, easing field extension.
% Flow matching is sequential, simpler to understand (Even tough there are plans
% to make it flexible).

\vspace{.1in}   
\noindent \textbf{Oflib.} This independent library converts OpenFlow messages in a network format to an internal format used by \ac{s} and vice-versa. The process of converting messages is known as pack and unpack. Packing/unpacking a message usually means to add/remove padding bits, but it can also involve the conversion of complex \ac{tlv} fields into the most appropriate data structure. One example is the case of the flow match fields, which are translated into hash maps for dynamic and fast access. The Oflib should be the starting point to anyone willing to extend the OpenFlow protocol with new messages.

\vspace{.1in}
\noindent \textbf{Packet Parser.} A Pipeline packet that comes from the switch ports has the header fields extracted by the Packet Parser first. The parsing is automated by the Netbee library~\cite{netbee}. Netbee uses a NetPDL~\cite{risso2006netpdl} database in the format of \ac{xml} that contains the definition of the packet headers supported by OpenFlow 1.3. The NetPDL approach has been a powerful component that eases the addition of new fields to OpenFlow, specially in the case of variable headers such as the IPv6 Extension headers~\cite{denicol2011ipv6}. 

\vspace{.1in}   
\noindent \textbf{Flow Tables.} They are the initial part of the OpenFlow pipeline. The fields of a packet's header, parsed by the Packet Parser, are matched against the flows installed in the Flow Tables of the software switch. Matched packets are subject to a set of instructions that may include actions over the packet, e.g., setting one of the fields, or further processing by another table of the Pipeline. The software switch default behavior is to drop any packet that does not match a flow. The current version of the software switch defines a number of 64 tables in the Pipeline, however, that value can be easily changed to accommodate more.

In \ac{s}, the Flow Tables perform matching in the simplest possible form. Flows are stored in priority order in a linked list. Thus, finding a matching entry has $\mathcal{O}(n)$ complexity. Flow Tables also maintain a list with references to flow entries with hard and idle timeouts, enabling faster checks of expired flows. 

\vspace{.1in}
\noindent \textbf{Group Table.} The Group Table enables different ways to forward packets. It can be used for fast-failover of ports, broadcast and multicast and even to implement Link Aggregation~(LAG). The software switch supports all the group entry types defined by the OpenFlow 1.3 specification. Actions in a group of the type Select are picked by a simple Round-Robin algorithm. Entries from the Group Table are stored in a hash map for $\mathcal{O}(1)$ retrieval.

\vspace{.1in}   
\noindent \textbf{Meter Table.} Metering was introduced in OpenFlow 1.2 and it gives the possibility to perform Quality of Service~(QoS) in a per flow basis. The software switch supports the two types available on OpenFlow 1.3, the simple Drop and the Differentiated Services Code Point (DSCP) remark. A basic Token Bucket algorithm is used to measure the per flow rate and decide if the Meter instruction should be applied or not.

\vspace{.1in}   
\noindent \textbf{Secure Channel.} The secure channel is a standalone program to set up a connection between the switch and a controller. The division from the datapath happens because OpenFlow does not define the connection method, so implementations are free to define the connection protocol; e.g: \ac{tcp} or \ac{ssl}; to establish connections. Although having $secure$ on its name, at the moment, the component supports only \ac{tcp} connections. Support for secure oriented protocols, such as \ac{ssl}, require updates to the Secure Channel code.

\vspace{.1in}   
\noindent \textbf{Dptcl.} The switch includes a command line tool to perform simple monitoring and administration tasks. With Dpctl one can modify and check the current state of switches. A few example of possible tasks: add new flows, retrieve current flow statistics and query the state of ports. 

\section{Selected Use Cases}
\label{sec:usecases}

This section presents a series of \ac{s} use cases in which some of the authors have contributed. The nature of these use cases is diverse and can be classified in four types: (1) extensions of the \ac{s} switch, (2) implementation of research ideas, (3) deployment of proof of concepts, and (4) research analysis or teaching SDN architectural concepts. Altogether, they showcase \ac{s} value in supporting industry, research, and academic institutions.

\subsection{BEBA}

\subsubsection{OpenState Extension:}

BEhavioural BAsed forwarding (BEBA)~\cite{BEBA} is a European H2020 project on SDN data plane programmability.
The BEBA software prototype has been built on top of \ac{s} with two main contributions: support for stateful packet forwarding, based on OpenState~\cite{Bianchi14}, and packet generation, based on InSPired (InSP) switches~\cite{Bifulco16}.

OpenState is an OpenFlow extension that allows implementing stateful applications in the data plane: the controller configures the switches to autonomously (i.e., without relying on the controller) and dynamically adapt the forwarding behavior. The provided abstraction is based on Finite State Machines where each state defines a forwarding policy and state transitions are triggered by packet-level and time-based events. \ac{s} has been extended using the OpenFlow experimenter framework and adding to each flow table an optional state table to keep track of flow states. Stateful forwarding is enabled thanks to the ability to match on flow state in the flow table and the availability of a data plane action to update the state directly in the fast path. Stateful processing is configured by the controller via experimenter OpenFlow messages.

InSP is an API to define in-switch packet generation operations, which include the specification of triggering conditions, packet format and related forwarding actions. An application example shows how the implementation of an in-switch ARP responder can be beneficial to both the switch and controller scalability.

The additional flexibility introduced by BEBA switches enables several use cases which get benefits from the reduced controller-switch signaling overhead regarding latency and processing. Cascone et. al~\cite{Cascone17} present an example application showing how BEBA allows implementing a programmable data plane mechanism for network resiliency which provides guaranteed failure detection and recovery delay regardless of controller availability.
StateSec~\cite{Boite17} is another example of stateful application combining the efficient local monitoring capabilities of BEBA switches with entropy-based algorithm running on the controller for DDoS Protection. 

\subsubsection{Performance enhancements:}

The second goal for BEBA has been the performance improvement of the data plane. To tackle such a problem, a major refactoring has been put on the field. 

The set of patches applied to the code base of \ac{s} comprises a Linux kernel bypass to improve the IO network performance, a new design for the packet handle data--type and the full exploitation of the multi-core architectures.

First, the native PF\_PACKET Linux socket originally utilized to send/receive packets has been replaced with libpcap~\cite{libpcap}. The aim of this refactoring is twofold: on the one hand, it makes the code more portable, on the other, it facilitates the integration with accelerated kernel-bypass already equipped with custom pcap libraries. 

Second, the structure of the packet-handle has been flattened into a single buffer to replace the multi-chunk design abused in the original code. This change permits to save a dozen of dynamic memory allocations (and related deallocations) on a per-forwarding basis, which represents a remarkable performance improvement per-se.

Finally, to tackle the parallelism of the multicore architecture, the PFQ~\cite{pfq} framework has been adopted. The reason for such a choice over more widely used solution like DPDK is the fine-grained control of the packet--distribution offered by PFQ off-the-shelf. The ability to dispatch packets to multiple forwarding processes, transparently and with dynamic degrees of flow-consistency, is fundamental to a stateful system like BEBA, where hard consistency guarantees are required by the XFSM programs loaded on the switch.

The remarkable acceleration obtained (nearly 100x) allows the prototype to full switch 4/5 Mpps per--core and to forward the 10G line rate of 64 bytes-long packets with four cores on our 3 GHz but old Xeon architecture.

A comprehensive description of the various techniques utilized in the BEBA switch, as well as the acceleration contribution of every single patch, are presented in~\cite{ofss-accel}.

% \begin{enumerate}
%   \item OpenState Extension: description + ofsoftswitch extension + use case 
%   \item Performance optimization: general patches + PFQ (kernel bypass)
%\end{enumerate}

\subsection{AOSS: OpenFlow hybrid switch}

AOSS~\cite{Alvarez-Horcajo17} emerged as a solution for the potential scalability problems of using SDN alone to control switch behavior. Its principle is to delegate part of the network intelligence \textit{back} to the network device --or switch--, thus resulting in a hybrid switch. Its implementation is based on the --currently-- most common \ac{sbi} protocol: OpenFlow. 

AOSS accepts proactive installation of OpenFlow rules in the switch and, at the same time, it is capable of forwarding packets through a shortest path when no rule is installed. To create shortest paths, it follows the locking algorithm of All-Path's switches~\cite{Rojas15}, which permits switches to create minimum latency paths on demand, avoiding loops without changing the standard Ethernet frame. 

An example of application for AOSS could be a network device that needs to drop some type of traffic (firewall), but forward the rest. In this case, the firewall rules would be installed proactively by the SDN controller and new packets arriving with no associated match would follow the minimum latency path to destination. This reduces drastically the control traffic, as the SDN controller just needs to bother about the proactive behavior and is not required to reply to \texttt{PACKET\_IN} messages, usually generated for any unmatched packet.

AOSS is particularly favorable for scenarios as the one described above, but its implementation still does not support composition of applications or reactive SDN behavior. Nevertheless, it is a good approach for hybrid environments where the network intelligence is not strictly centralized, thus improving overall performance.

\subsubsection{AOSS Implementation:}
\begin{figure}[!ht]
    \center
    \includegraphics[width=0.6\textwidth]{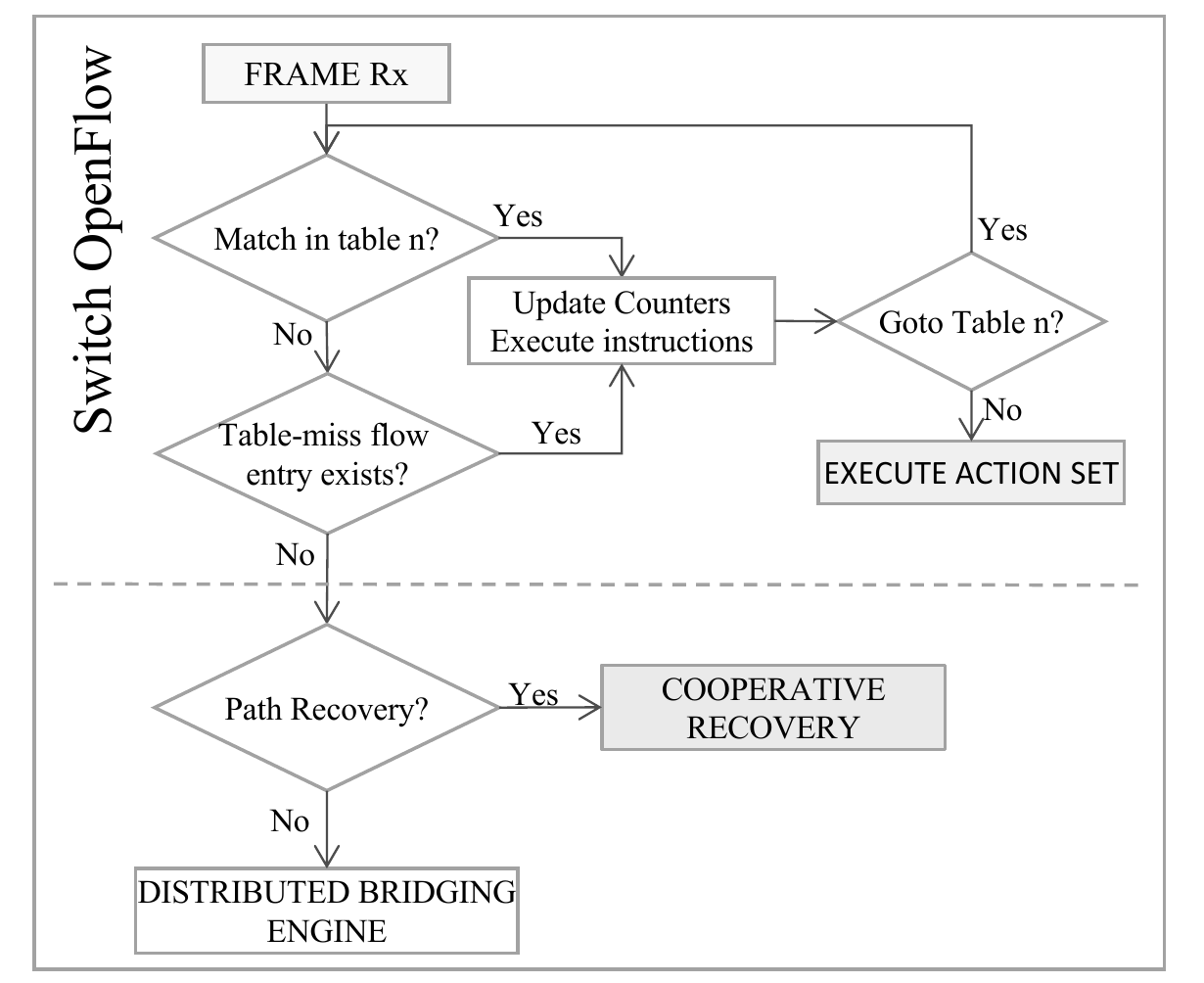}
    \caption{AOSS's Frame Processing \cite{Alvarez-Horcajo17}}
    \label{fig:AOSS_Frame}
\end{figure}

To create a \ac{poc} of AOSS, different open-source SDN software switches were analyzed. Although \ac{ovs} was first in the list, due to its kernel-based (and thus higher performance) nature, leveraging its code to quickly build a PoC was laborious. Therefore, the code of \ac{s} was adopted instead. 
%Ofsoftswitch13 implements the characteristics defined by OpenFlow 1.3 protocol for a SDN-OpenFlow switch. Starting from this definition, 
AOSS needs some modifications to generate the hybrid system. The main one requires inserting an autonomous path selection for all packets with no associated match in the OpenFlow table. Fig.~\ref{fig:AOSS_Frame} reflects these functional changes.

\begin{figure}[!ht]
    \center
    \includegraphics[width=0.5\textwidth]{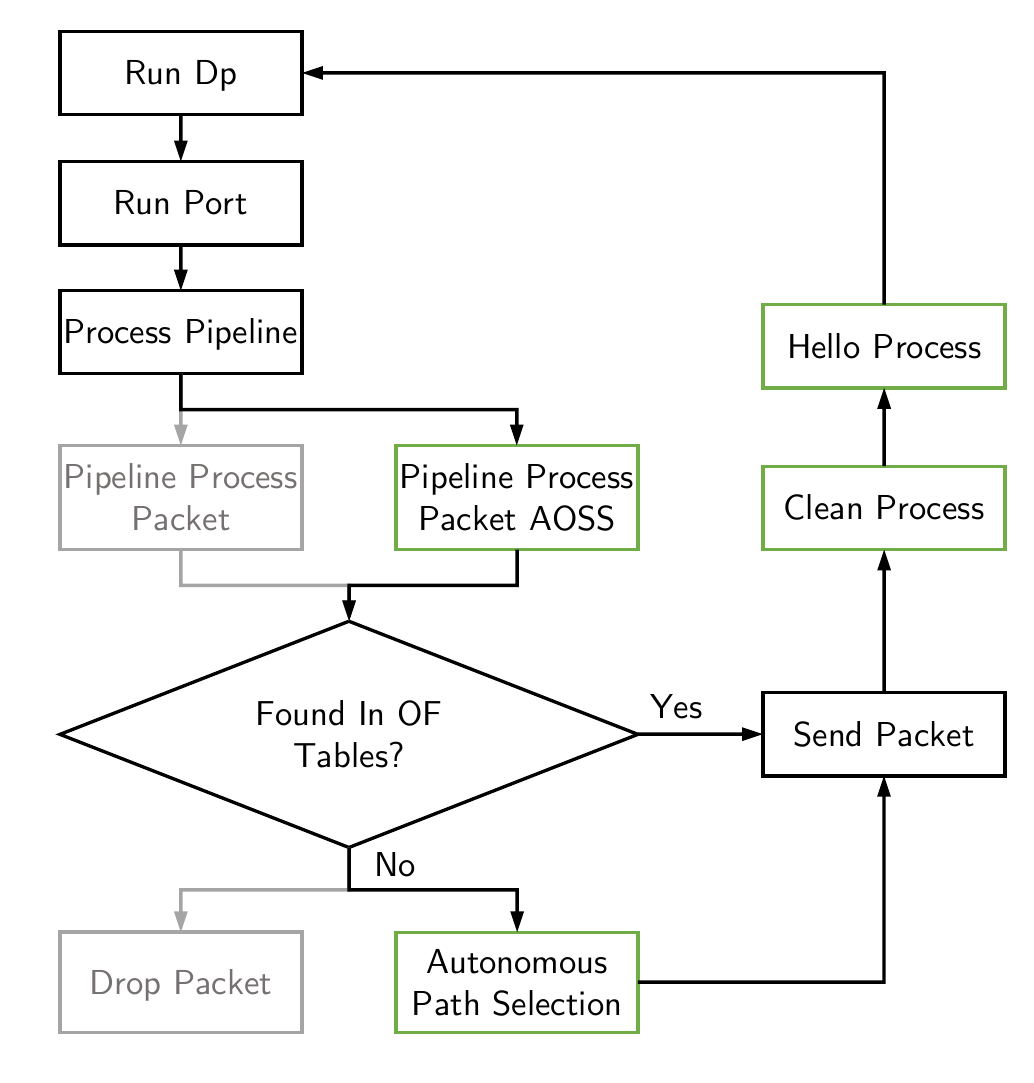}
    \caption{AOSS's Functional Flow Chart}
    \label{fig:aoss_functional_flow_chart}
\end{figure}

Regarding AOSS implementation, two functional changes and two new functions are defined, as defined in Fig.~\ref{fig:aoss_functional_flow_chart}. The first change is a modification in the \textit{Pipeline Process Packet Function} to guarantee compatibility with the autonomous path selection protocol. The second change modifies the drop packet function to create the minimum latency path. As for the new functions, the first is responsible for cleaning the new forwarding tables and the second sends special control frames to allow path recovery after a network failure.

\subsection{OnLife: Deploying the CORD project in a national operator}
OnLife~\cite{Montero17} is a deployment of the CORD project~\cite{Peterson16} in Telefonica's\footnote{Main Spanish telecommunications provider} central offices. The main purpose of OnLife is to bring services as closer to the final user as possible, to enhance their quality, and its first principle is to create a clean network deployment from scratch, with no legacy protocols (e.g. allowing only IPv6 and avoiding IPv4).

The first step in OnLife was building a PoC, purely software-based, to prove its foundations. In CORD, some of the applications in the SDN framework (namely ONOS~\cite{Berde14}) require IEEE 802.1ad QinQ tunneling~\cite{qinq} to classify different flows of traffic inside the data center. Therefore \ac{s} was leveraged as \ac{ovs} does not support this feature.

\ac{s} allowed the initial design of the project, although some initial incompatibilities were found in the communication between ONOS and the switches, solved afterwards. The main conclusion is that \ac{s} became a crucial piece for these deployments, and specific efforts should be made to increase its visibility and community support.

\subsection{\ac{s} as a teaching resource}
One of the first degrees that teaches the SDN and NFV technologies as tools for the emerging communication networks, specifically 5G networks, is the Master in NFV and SDN for 5G Networks of the University Carlos III of Madrid~\cite{mastersdnnfv}. 

\ac{s} is part of the syllabus, presented together with \ac{ovs}, as one of the two main open source software SDN switches. As its main feature, its easy customization is highlighted. 
\section{Fostering Research \& Standardization } %(through open source)
\label{sec:research}

Following the classification provided in the previous use cases, this section is devoted to create a brief catalog of the different works found in the literature that have leveraged \ac{s}. The categories are: research implementations or evaluations, \ac{poc} implementations, and \ac{sdn} switch comparatives, and teaching resources. The resulting grouping is summarized in Table~\ref{table:classification}.

\begin{table*}%[htbp]
\begin{adjustbox}{max width=\textwidth}
\begin{threeparttable}
\caption{Classification of works that leverage  \ac{s}\label{table:classification}}
\begin{tabular}{|c|c|c|c|c|}
\hline
\backslashbox{\textbf{Article}}{\textbf{Properties}}& \textbf{Description} & \textbf{Type} & \textbf{Why?}  \\
\hline
\hline

%Research Implementation
\textbf{OpenState~\cite{Bianchi14}}& 
\begin{tabular}{@{}c@{}}OpenFlow extension for\\stateful applications\end{tabular} 
& Research implementation 
& Pipeline modification  \\
\hline
%-----
\textbf{InSP~\cite{Bifulco16}}& 
\begin{tabular}{@{}c@{}}API to define in-switch\\packet generation operations\end{tabular} 
& Research implementation 
& Pipeline modification   \\
\hline
%-----
\textbf{AOSS~\cite{Alvarez-Horcajo17} }& 
\begin{tabular}{@{}c@{}}Stateful (hybrid)\\\ac{sdn} switch\end{tabular} 
& Research implementation 
& Pipeline modification   \\
\hline
%-----
\textbf{OPP~\cite{Bianchi16}}& 
\begin{tabular}{@{}c@{}}Platform-independent stateful\\in-network processing\end{tabular} 
& Research implementation 
& Pipeline modification   \\ 
\hline
%-----
\textbf{BPFabric~\cite{Jouet15,Jouet17}}& 
\begin{tabular}{@{}c@{}}On-the-fly data plane packet processing pipeline\\and direct manipulation of network state\end{tabular} 
& Research implementation 
& Pipeline modification  \\
\hline
%-----
\textbf{Fast switchover/failover~\cite{Nguyen14}}& 
\begin{tabular}{@{}c@{}}New switchover method based on \\active/active mode (select group)\end{tabular} 
& Research implementation 
& Pipeline modification   \\
\hline
%-----
\textbf{FlowConvertor~\cite{Pan17}}& 
\begin{tabular}{@{}c@{}}Algorithm that provides\\portability across switch models\end{tabular} 
& Research implementation 
& Pipeline modification   \\ 
\hline
%-----
\textbf{Chronus~\cite{Zheng17}}& 
\begin{tabular}{@{}c@{}}Scheduled consistent\\network updates\end{tabular} 
& Research implementation 
& Pipeline modification   \\ 
\hline
%-----
\textbf{REV~\cite{Zhang17}}& 
\begin{tabular}{@{}c@{}}New security primitive\\for SDN\end{tabular} & Research implementation 
& Pipeline modification   \\ 
\hline
%-----
\textbf{RouteFlow~\cite{vidal122013building}}& 
\begin{tabular}{@{}c@{}} OpenFlow 1.x Dataplane for virtual \\ routing services \end{tabular} 
& Research implementation 
& \begin{tabular}{@{}c@{}} OpenFlow version interoperability \\ 
and Group Tables \end{tabular}  \\
\hline
%-----
\textbf{\begin{tabular}{@{}c@{}}TCP connection\\handover~\cite{Binder15}\end{tabular} }& 
\begin{tabular}{@{}c@{}}New method of TCP connection\\handover in SDN\end{tabular} 
& Research implementation 
& Modification of OpenFlow 1.3  \\
\hline
%-----
\textbf{\begin{tabular}{@{}c@{}}Facilitating ICN\\with SDN~\cite{Zuraniewski17}\end{tabular} }& 
\begin{tabular}{@{}c@{}}Leveraging SDN\\for ICN scenarios\end{tabular} 
& Research implementation 
& Extension of OpenFlow  \\
\hline
%-----
\textbf{\AE therFlow~\cite{Yan15}}& 
\begin{tabular}{@{}c@{}}Application of SDN principles\\to wireless networks\end{tabular} 
& Research implementation 
& Extension of OpenFlow  \\
\hline
%-----
\textbf{CrossFlow~\cite{Shome15},\cite{Shome17}}& 
\begin{tabular}{@{}c@{}}Application of SDN principles\\to wireless networks\end{tabular} 
& Research implementation 
& Extension of OpenFlow   \\
\hline
%-----
\textbf{\begin{tabular}{@{}c@{}}Media Independent\\Management~\cite{Guimaraes14}\end{tabular} }& 
\begin{tabular}{@{}c@{}}Dynamic link information acquisition\\to optimize networks\end{tabular} 
& Research implementation 
& Extension of OpenFlow  \\
\hline
%-----
\textbf{\begin{tabular}{@{}c@{}}Automatic failure\\recovery~\cite{Kuzniar13}\end{tabular} }& 
\begin{tabular}{@{}c@{}}Proxy between SDN controller\\and switches to handle failures\end{tabular} 
& Research implementation 
& Reuses oflib from ofsoftswitch13 \\
\hline
%-----
\textbf{OFSwitch13~\cite{Chaves16}}& 
\begin{tabular}{@{}c@{}}Module to enhance the ns-3 simulator\\ with OpenFlow 1.3\end{tabular} 
& Research implementation 
& Reuses ofsoftswitch13  \\
\hline
%-----
\textbf{Time4~\cite{Mizrahi16}}& 
\begin{tabular}{@{}c@{}}Approach for network updates\\(adopted in OpenFlow 1.5)\end{tabular} 
& Research implementation 
& \texttt{Bundle} feature   \\ 
\hline
%-----
\textbf{OFLoad~\cite{Trestian17}}& 
\begin{tabular}{@{}c@{}}OF-Based Dynamic Load Balancing \\for data center networks\end{tabular} 
& Research implementation 
& OpenFlow \texttt{group} option  \\ 
\hline
%-----
\textbf{\begin{tabular}{@{}c@{}}Blind Packet Forwarding\\in hierarchical architecture~\cite{Simsek14}\end{tabular} }& 
\begin{tabular}{@{}c@{}}Implementation of the\\extended BPF\end{tabular} 
& Research implementation 
& N/D  \\ %TODO
\hline
%-----
\textbf{GPON SDN Switch~\cite{Lee16}}& 
\begin{tabular}{@{}c@{}}GPON based OpenFlow-enabled \\SDN virtual switch\end{tabular} 
& Research implementation 
& Part of the architecture  \\ 
\hline
%-----
\textbf{\begin{tabular}{@{}c@{}}Traffic classification\\with stateful SDN~\cite{Sanvito17}\end{tabular}}& 
\begin{tabular}{@{}c@{}}Traffic classification in the data plane\\to offload the control plane \end{tabular} 
& Research implementation 
& \begin{tabular}{@{}c@{}}Leverages OpenState~\cite{Bianchi14} \\and OPP~\cite{Bianchi16} \end{tabular}  \\ 
\hline
%-----
\textbf{\begin{tabular}{@{}c@{}}Traffic classification and control\\with stateful SDN~\cite{Bianco17}\end{tabular}}& 
\begin{tabular}{@{}c@{}}Traffic classification in the data plane\\to offload the control plane \end{tabular} 
& Research implementation 
& Leverages OpenState~\cite{Bianchi14}   \\ 
\hline
%-----
\textbf{SPIDER~\cite{Cascone17}}& 
\begin{tabular}{@{}c@{}}OpenFlow-like pipeline design for failure\\detection and fast reroute of traffic flows \end{tabular} 
& Research implementation 
& Leverages OpenState~\cite{Bianchi14}   \\ 
\hline
%-----
\textbf{StateSec~\cite{Boite17}}& 
\begin{tabular}{@{}c@{}}In-switch processing capabilities\\to detect and mitigate DDoS attacks \end{tabular} 
& Research implementation 
& Leverages OpenState~\cite{Bianchi14}   \\ 
\hline
%-----
\textbf{\begin{tabular}{@{}c@{}}Load balancers\\evaluation~\cite{Silva17}\end{tabular} }& 
\begin{tabular}{@{}c@{}}Evaluation of different\\load balancer apps\end{tabular} 
& Research evaluation 
& Leverages OpenState~\cite{Bianchi14}  \\
\hline
%-----
\textbf{\begin{tabular}{@{}c@{}}Recovery of multiple \\failures in SDN~\cite{Zahid17}\end{tabular} }& 
\begin{tabular}{@{}c@{}}Comparison of OpenState and OpenFlow\\in multiple-failure scenarios\end{tabular} 
& Research evaluation 
& Leverages OpenState~\cite{Bianchi14}  \\
\hline
%-----

%PoC Implementation
\textbf{UnifyCore~\cite{Nagy15}}& 
\begin{tabular}{@{}c@{}}Mobile architecture implementation in which\\ofsoftswitch13 is leveraged as a fordwarder\end{tabular} 
& PoC implementation 
& MAC tunneling  \\
\hline
%-----
\textbf{ADN~\cite{Tegueu16}}& 
\begin{tabular}{@{}c@{}}Architecture that provides QoS \\on an application flow basis\end{tabular} 
& PoC implementation 
& \begin{tabular}{@{}c@{}}Full support of OpenFlow 1.3\\(meters and groups \texttt{all/select})\end{tabular}  \\
\hline
%-----
\textbf{\begin{tabular}{@{}c@{}}TCP connection handover\\for hybrid honeypot systems~\cite{Fan17}\end{tabular}}& 
\begin{tabular}{@{}c@{}}TCP connection handover mechanism \\implemented in SDN\end{tabular}
& PoC implementation
& Data plane programmability   \\ 
\hline
%-----
\textbf{\begin{tabular}{@{}c@{}}Multiple Auxiliary TCP/UDP\\Connections in SDN~\cite{Yang17}\end{tabular}}& 
\begin{tabular}{@{}c@{}}Analysis and implementation \\of multiple connections in SDN\end{tabular}
& PoC implementation
& Extension of OFSwitch13~\cite{Chaves16}  \\ 
\hline
%-----
\textbf{\begin{tabular}{@{}c@{}}State-based security protection\\mechanisms in SDN~\cite{Arumugam17}\end{tabular}}& 
\begin{tabular}{@{}c@{}}Demonstration of the\\SDN Configuration (CFG)  protection\end{tabular}
& PoC implementation
& Leverages OpenState~\cite{Bianchi14}  \\ 
\hline
%-----
\textbf{\begin{tabular}{@{}c@{}}Advanced network\\functions~\cite{Bonola17}\end{tabular}}& 
Stateful data-plane network functions
& PoC implementation 
& Leverages OPP~\cite{Bianchi16}  \\ %TODO
\hline
%-----
\textbf{PathMon~\cite{Wang16}}& 
Granular traffic monitoring
& PoC implementation 
& N/D  \\ %TODO
\hline
%-----
\textbf{\begin{tabular}{@{}c@{}}QoT Estimator in SDN-Controlled\\ROADM networks~\cite{Diaz-Montiel18}\end{tabular}}& 
\begin{tabular}{@{}c@{}}Implementation of a QoT estimator\\in a simulated optical network\end{tabular}
& PoC implementation
& N/D  \\ 
\hline
%-----
\textbf{OPEN PON~\cite{Silva16}}& 
\begin{tabular}{@{}c@{}}Integration of 5G core \\and optical access networks\end{tabular}
& \begin{tabular}{@{}c@{}}PoC implementation\\ (MSc Thesis)\end{tabular}
& \begin{tabular}{@{}c@{}}Support of IEEE 1904.1 SIEPON,\\meters and Q-in-Q\end{tabular}  \\ 
\hline
%-----
\textbf{\begin{tabular}{@{}c@{}}Stochastic Switching\\Using OpenFlow~\cite{Shahmir13}\end{tabular}}& 
\begin{tabular}{@{}c@{}}Analysis and implementation \\of stochastic routing in SDN\end{tabular}
& \begin{tabular}{@{}c@{}}PoC implementation\\ (MSc Thesis)\end{tabular}
& \begin{tabular}{@{}c@{}}\texttt{Select} function of\\\texttt{Group} feature\end{tabular}   \\ 
\hline
%-----

%SDN comparative
\textbf{OpenFlow forwarders~\cite{Sulak16}}& 
\begin{tabular}{@{}c@{}}Routing granularity\\ in OpenFlow 1.0 and 1.3\end{tabular} 
& SDN switch comparative 
& N/A  \\
\hline
%-----
\textbf{Open source SDN~\cite{Tantayakul17}}& 
\begin{tabular}{@{}c@{}}Performance of open source \\SDN virtual switches\end{tabular} 
& SDN switch comparative 
& N/A   \\
\hline
%-----

%Teaching resource
\textbf{Visual system to learn OF~\cite{Fujita17}}& 
\begin{tabular}{@{}c@{}}A visual system to support learning\\ of OpenFlow-based networks\end{tabular} 
& Teaching resource
& N/D \\
\hline
%-----

\end{tabular}
    \begin{tablenotes}
      \small
      \item N/A means \emph{not applicable}.
      \item N/D means \emph{not defined}.
    \end{tablenotes}
\end{threeparttable}
\end{adjustbox}
\end{table*}

%Table: https://docs.google.com/spreadsheets/d/1XUDUUAXtNuLvYvxWQsC94guMLg9O8UghbisxE5JGMhE
%Elisa: I added more articles from a new search from July 2018.

\subsection{Research implementations or evaluations}
Three research implementations have already been introduced in the use cases, namely \textbf{OpenState}~\cite{Bianchi14}, \textbf{InSP}~\cite{Bifulco16} and \textbf{AOSS}~\cite{Alvarez-Horcajo17}. All of them envision alternative architectures for \ac{sdn} in which network switches recover part of the intelligence of the network and, accordingly, they leverage \ac{s} thanks to its easily modifiable pipeline.

Also based on pipeline modifications, \textbf{Open Packet Processor (OPP)}~\cite{Bianchi16} enhances the approach of OpenState to support extended Finite State Machines, which broadens the potential functionality of the data plane. 
\textbf{BPFabric}~\cite{Jouet15,Jouet17} defines an architecture that allows instantiating and querying, on-the-fly, the packet processing pipeline in the data plane. 

Regarding the evolution of current \ac{sbi} protocols (namely OpenFlow), an alternative switchover procedure (active/active instead of active/standby) is presented in~\cite{Nguyen14}, which leverages the \texttt{select} group of \ac{s}. 
\textbf{RouteFlow}~\cite{vidal122013building} is a pioneering architectural proposal to deliver flexible (virtual) IP routing services over OpenFlow networks~\cite{routeflow-hotsdn12} (developed by the same core research group at CPqD behind \ac{s}), which extensively used the software switch for fast prototyping, interoperability tests with OpenFlow 1.2 and 1.3, and new features such as group tables.

Considering the heterogeneity of switch pipeline implementations, \textbf{FlowConvertor}~\cite{Pan17} defines an algorithm that provides portability across different models. To prove the idea, it applies it to a \ac{s} switch, as it demonstrates to have a flexible and programmable pipeline. 
Another research topic in relation to the \ac{sbi} are transactional operations and consistent network updates (currently OpenFlow does not support these types of procedures), and \textbf{Chronus}~\cite{Zheng17} modifies \ac{s} to provide scheduled network updates, to avoid potential problems, such as communication loops or blackholes. 
Finally, \textbf{REV}~\cite{Zhang17} designs a new security primitive for \ac{sdn}, specifically aimed to prevent rule modification attacks.

In the specific case of enhancements of OpenFlow, an extension of OpenFlow 1.3 thanks to \ac{s} is introduced in~\cite{Binder15}, which includes two new actions (\texttt{SET\_TCP\_ACK} and \texttt{SET\_TCP\_SEQ}) to modify the ACK and SEQ values in TCP connections. 
Alternatively, the matching capabilities of OpenFlow have been extended in~\cite{Zuraniewski17} to provide an optimal parsing of packets in the context of \ac{icn}.  
Both \textbf{\AE therFlow}~\cite{Yan15} and \textbf{CrossFlow}~\cite{Shome15} study how to evolve OpenFlow to include the \ac{sdn} principles in wireless networks. In this regard, \ac{s} acts as an OpenFlow agent with custom extensions.
Another extension of OpenFlow is provided in~\cite{Guimaraes14}, were the authors design a framework where the key is media independent management. 

Different research implementations are based on \ac{s} because they simply wanted to leverage some piece of its code. 
For example, the automatic failure mechanism described in~\cite{Kuzniar13} reuses the \texttt{oflib} library. 
\textbf{OFSwitch13}~\cite{Chaves16} reuses the whole code of \ac{s} to incorporate the support of OpenFlow 1.3 in the network simulator ns-3. 
\textbf{Time4}~\cite{Mizrahi16} reuses the \texttt{bundle} feature to implement an approach for network updates (actually adopted in OpenFlow 1.5). 
\textbf{OFLoad} \cite{Trestian17} leverages the OpenFlow \texttt{group} option from \ac{s} to design an strategy for dynamic load balancing in \ac{sdn}. 
The principles of Blind Packet Forwarding (BPF) also reuse the code of \ac{s} for the implementation. 
A textbf{GPON SDN Switch}, where \ac{s} is part of the architecture, is also designed and developed in~\cite{Lee16}. 

Finally, several research ideas leverage OpenState and, thus, \ac{s}. 
The first two were already mentioned previously: \textbf{SPIDER}~\cite{Cascone17} and \textbf{StateSec}~\cite{Boite17}, both examples of stateful applications aimed to provide enhanced network resiliency and monitoring, respectively. Also, \textbf{traffic classificators} based on OpenState are also presented in~\cite{Sanvito17} and ~\cite{Bianco17}.
Additionally, an evaluation of \ac{sdn} load balancing implementations is performed in~\cite{Silva17}, and authors in~\cite{Zahid17} compare recovery of \ac{sdn} from multiple failures for OpenFlow vs. OpenState. 

\subsection{\ac{poc} implementations}
\ac{s} has also been part of different \ac{poc} implementations. 
For example, \textbf{UnifyCore}~\cite{Nagy15} is an integrated mobile network architecture, based on OpenFlow but leveraging legacy infrastructure. They evaluate the MAC tunneling implemented in \ac{s} with \texttt{iperf}.
\textbf{ADN}~\cite{Tegueu16} describes an architecture that provides QoS based on application flow information, and they chose \ac{s} because it fully supports OpenFlow 1.3. 
Authors in~\cite{Fan17} implemented a novel TCP connection handover mechanism with \ac{s}, aimed to provide transparency to honeypots by generating the appropriate sequence and acknowledgement numbers for the TCP redirection mechanism to work.

One \ac{poc} leveraged OFSwitch13 (\ac{s} in ns-3) to support multiple transport connections in \ac{sdn} simulations~\cite{Yang17}, while authors in~\cite{Arumugam17} leverage OpenState to demonstrate that stateful data-plane designs can provide additional security for operations such as link reconfiguration or switch identification. Advanced network functions based on OPP are implemented and tested in~\cite{Bonola17}.

Out of curiosity, there are some works that use \ac{s} just as the \ac{sdn} software switch for no particular reason (as many others use \ac{ovs} by default). 
One of them is \textbf{PathMon}~\cite{Wang16}, which provides granular traffic monitoring. 
Another one is a QoT estimator for ROADM networks implemented and evaluated in~\cite{Diaz-Montiel18}.

Finally, two MSc. Thesis have also be developed based on \ac{s}. 
The first one is \textbf{OPEN PON}~\cite{Silva16}, which analyzes the integration between the 5G core and optical access networks. \ac{s} was selected because of different reasons, but mainly because of its support of standards, such as Q-in-Q (required to emulate the behaviour of the OLT modules), which is not properly implemented in OVS. 
The second one describes stochastic switching using OpenFlow~\cite{Shahmir13} and \ac{s} was once again chosen due to its good support of specific features, such as the \texttt{select} function.

\subsection{Comparative reports and Teaching resources}
In this last category, it is worth mentioning two comparison studies: a performance analysis of OpenFlow forwarders based on routing granularity~\cite{Sulak16}, and an experimental analysis of different pieces of software in an SDN open source environment~\cite{Tantayakul17}. The former compares \ac{s} with other switches, while the latter analyzes the role of \ac{s} in a practical \ac{sdn} framework. 
Finally, a nice teaching resource is described in~\cite{Fujita17}, where the authors present a system they put in practice to learn the basics of OpenFlow in a visual manner.

\section{Evaluation}
\label{sec:evaluation}

As previously stated, there are currently two main types of software switches for \ac{sdn} environments: \ac{ovs} and \ac{s}. The main conclusion is that \ac{ovs} performs much better, but it is hard to modify, while \ac{s} is particularly suitable for customizations and research work, even though its throughput limitations. This is just a qualitative comparison. 

For this reason, in this section, we provide an additional quantitative evaluation for \ac{ovs} vs. \ac{s}. More specifically, we will compare \ac{ovs} with the two main \textit{flavours} of \ac{s}, namely the \textbf{original} BOFUSS~\cite{ofsoftswitch13} and the \textbf{enhanced} version implemented by the \textit{BEBA}~\cite{BEBA-github} project. The comparison will be performed via two tests:
\begin{enumerate}
    \item Individual benchmarking of the three switches via iPerf~\cite{iperf}
    \item Evaluation in a data center scenario with characterized traffic and three different networks comprised of the different types of switches
\end{enumerate}
The main purpose is to provide a glance at the performance of \ac{s}, which might be good enough for many research scenarios, even if \ac{ovs} exhibits better results overall\footnote{A comparison of \ac{ovs} with other software switches, but without including \ac{s}, is provided in ~\cite{Fang18}.}.

\subsection{Individual benchmarking}
%- Throughput
For this first test, we directly benchmarked each of the three switches (\ac{ovs} and the two flavours of \ac{s}) with iPerf~\cite{iperf}. 
Our hardware infrastructure consisted of 1 computer powered by Intel(R) Core(TM) i7 processors (3,4 GHz) with 24 GB of RAM and Ubuntu 14.04 as Operating System. 
We deployed one single switch of each type and run iPerf 10 times for each scenario, obtaining the average throughput and standard deviation.

\begin{table}[t]
\centering
\resizebox{0.70\columnwidth}{!}{ 
    \begin{tabular}{ccc}
    \hline
    \textbf{Switch} & \textbf{$\overline{x}$} & \textbf{$\sigma$} \\ \hline \hline
    \textbf{OVS} & 51,413 Gbps&  2,6784 \\ \hline
    \textbf{Enhanced \ac{s}} & 1,184 Gbps& $3,945*10^{-3}$  \\ \hline
    \textbf{Original \ac{s}} & 0,186 Gbps& $6,86*10^{-5}$ \\ \hline
    \end{tabular}
}
\caption{Throughput of the three individual types of software switches, measured with iPerf} 
\label{tbl:results-one-switch}
\end{table}

The results are shown in Table~\ref{tbl:results-one-switch}. Although \ac{ovs} outperforms \ac{s}, it is important to notice how the enhanced switch surpass 1 Gbps,a result considered a reasonable throughput for most common networking scenarios.

\subsection{Evaluation in a data center scenario}
\begin{figure}[th!]
    \centering
    \includegraphics[width=0.7\columnwidth]{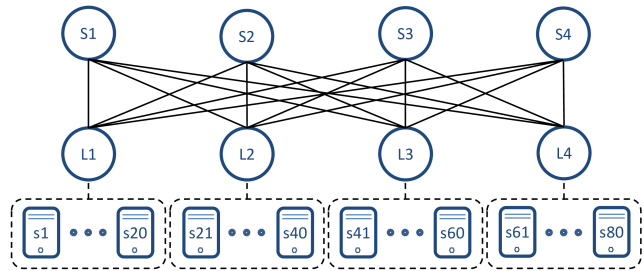}
    \caption{Spine-Leaf 4-4-20 evaluation topology~\cite{Alvarez-Horcajo17b}}
    \label{fig:SL44}
\end{figure}

%- Throughput and Flow Completion Time
For this second test, we focused on realistic scenarios data center deployments, where software switches could an essential part of the network infrastructure. 
We built a \textit{Spine-Leaf} topology~\cite{alizadeh_conga:_2014,he_presto:_2015,alizadeh_pfabric:_2013}, typically deployed for data center networks. More specifically, a 4-4-20 Spine-Leaf with 2 rows of 4 switches (4 of type \textit{spine} and 4 of type \textit{leaf}) and 20 servers per leaf switch for a total of 80 servers, as illustrated in Fig.~\ref{fig:SL44}.

\begin{figure}[th!]
    \includegraphics[width=0.9\columnwidth]{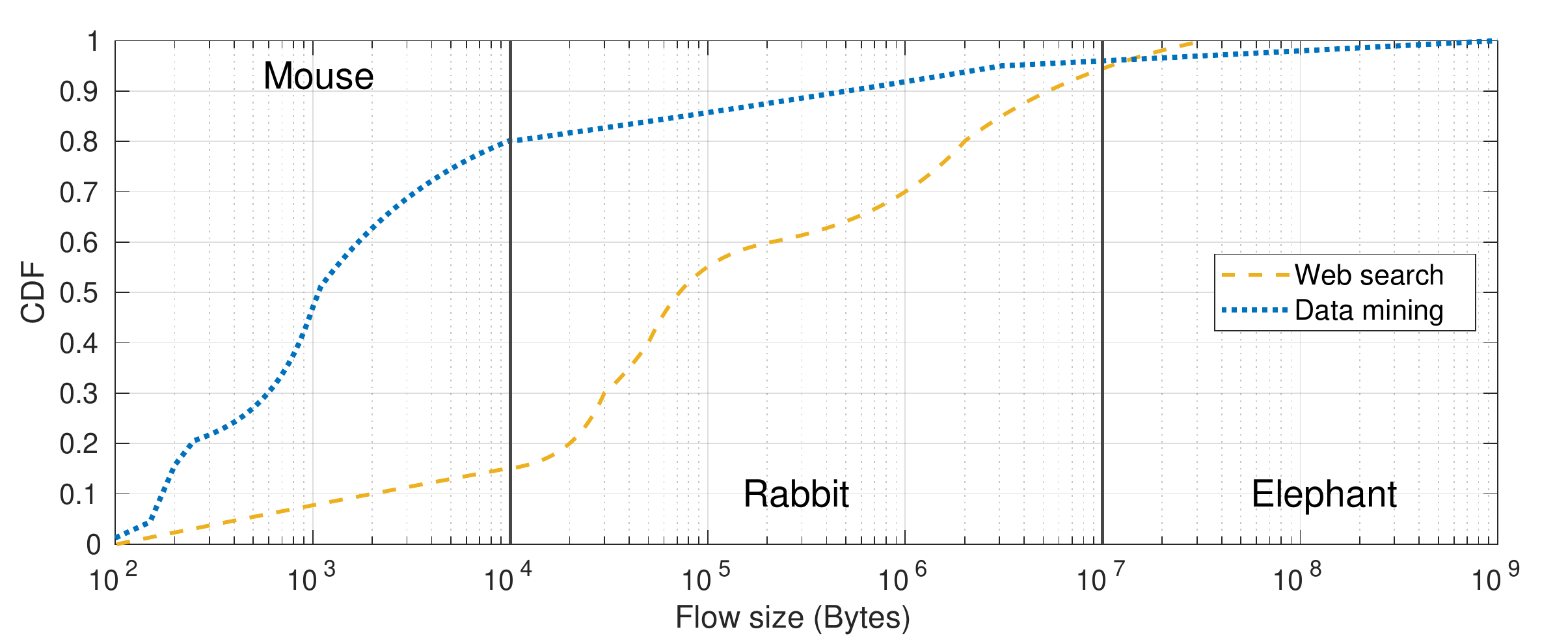}
    \caption{Flow size distributions~\cite{Alvarez-Horcajo17b}}
    \label{fig:Flow_Model}
\end{figure}
 
\begin{table}[th!]
\caption{Experimental setup of the data center scenarios}
\label{tab:experiments}       % Give a unique label
    \resizebox{0.70\columnwidth}{!}{ 
        \begin{tabular}{ll}
            \hline\noalign{\smallskip}
                Parameter & Value \\
                \noalign{\smallskip}\hline\noalign{\smallskip}
                Network topology & Spine-Leaf (4 - 4)\cite{alizadeh_conga:_2014} \\
                Servers per leaf switch & 20 \\
                Flow distribution  & Random inter-leaf \\
                Flow size distributions & Web search\cite{alizadeh_data_2010} \& Data mining \cite{greenberg_vl2:_2009} \\
                Network offered load (\%) & 10, 20 \& 40\% \\
                Link speed (Mpbs) & 100Mbps \\
                Run length (s) & 1800 s  \\
                Warm up time (s) & 800 s \\
                Number of runs & 10      \\
            \noalign{\smallskip}\hline
        \end{tabular}
    }
\end{table}

\begin{figure}[th!]
    \includegraphics[width=0.85\columnwidth]{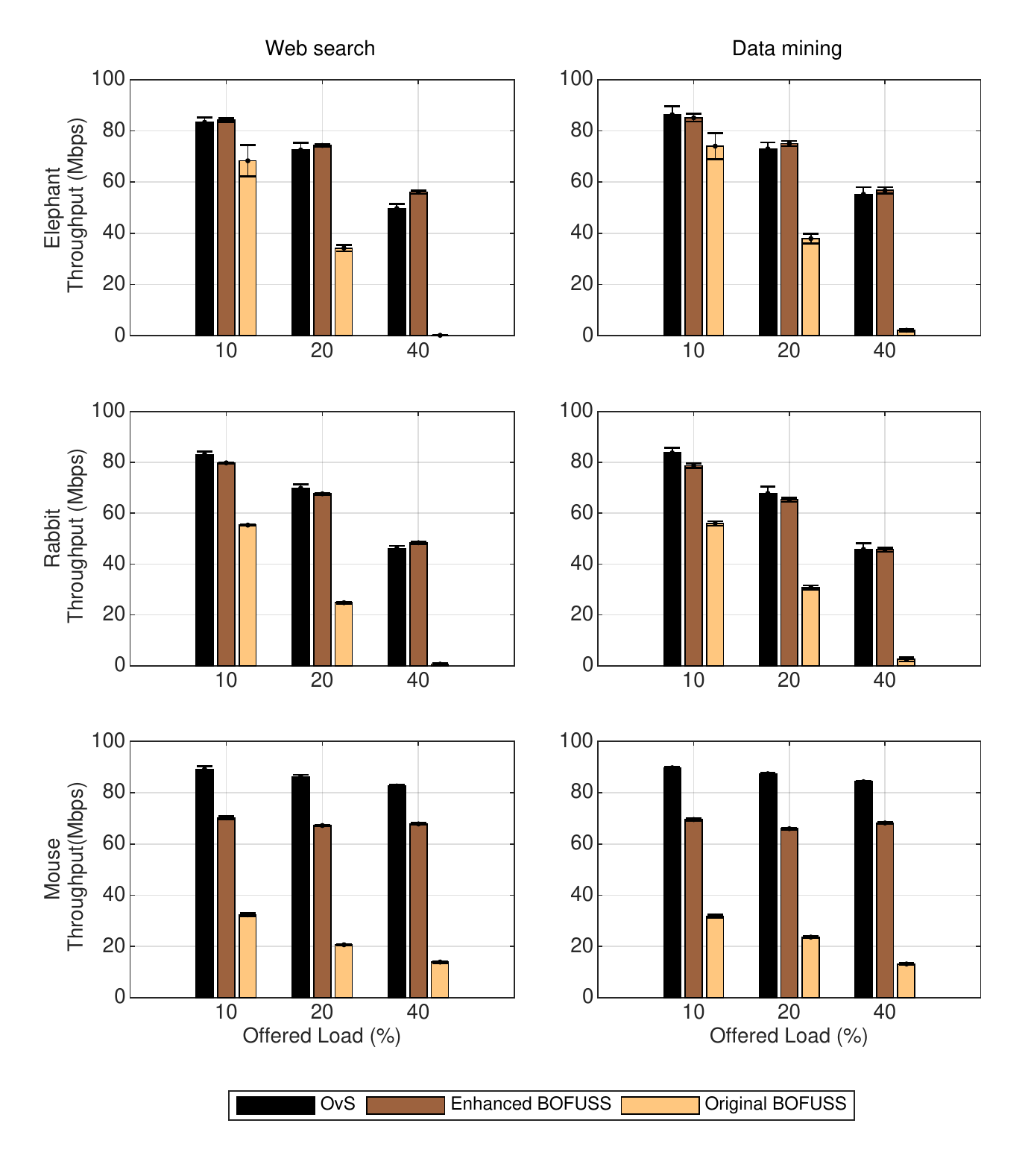}
    \caption{Throughput in the Spine-Leaf topology for each switch type}
    \label{fig:BW}
\end{figure}

\begin{figure}[th!]
    \includegraphics[width=0.85\columnwidth]{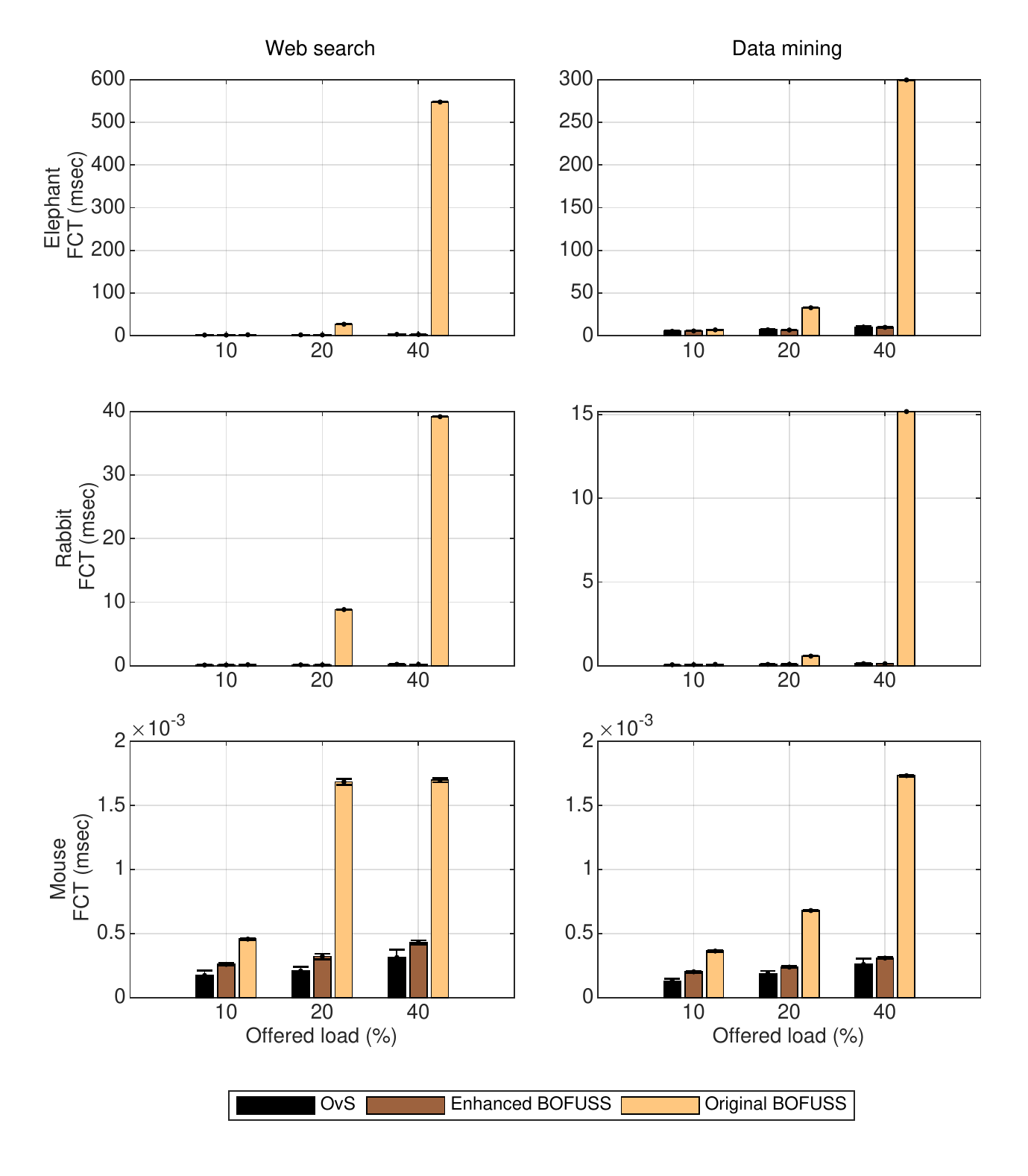}
    \caption{Flow Completion Time in the Spine-Leaf topology for each switch type}
    \label{fig:FCT}
\end{figure}

To emulate data center-like traffic, we developed a customized traffic generator~\cite{Alvarez-Horcajo17b}. 
This generator implements two different flow size distributions, namely \textit{Data Mining} and \textit{Web Search}, derived from experimental traces taken from actual data center networks~\cite{alizadeh_data_2010,greenberg_vl2:_2009}. Figure~\ref{fig:Flow_Model} shows the cumulative distribution function (CDF) of both distributions and also illustrates how flows are classified according to their size. 
Flows with less than 10 KB and more than 10 MB of data are considered \textit{mouse} and \textit{elephant} flows, respectively, as explained in \cite{greenberg_vl2:_2009}. The remaining flows are identified as \textit{rabbit} flows. 
Traffic flows are randomly distributed between any pair of servers attached to two different leaf switches with no further restrictions. 

Our hardware infrastructure consisted of a cluster of 5 computers powered by Intel(R) Core(TM) i7 processors (4,0 GHz) with 24 GB of RAM and Ubuntu 14.04 as Operating System, all of which are interconnected via a GbE Netgear GS116 switch. 
Each experiment was executed for 1800 seconds and repeated 10 times to compute 95\% confidence intervals. Additionally, we considered a warm-up time of 800 seconds to mitigate any transitory effect on the results. Table \ref{tab:experiments} summarizes the full setup of the conducted experiments.

To evaluate the performance of \ac{ovs} and the two flavours of \ac{s}, we measured throughput and flow completion time, which are depicted in Fig.~\ref{fig:BW} and Fig.~\ref{fig:FCT}, respectively\footnote{Raw evaluation data can be found at~\cite{bofuss-data}.}.  
The graphs are divided into the three types of flows, and we evaluated an increasing network offered load of 10\%, 20\% and 40\%.
The results show that \ac{ovs} and the enhanced \ac{s} perform quite similarly. In fact, they provide almost the same results for the elephants and rabbit flows (even more favorable for the enhanced \ac{s} in some cases), and better for \ac{ovs} in the case of the mouse flows. 
In all cases, the original \ac{s} is outperformed by \ac{ovs} and the enhanced \ac{s}. In fact, when the offered load reaches the 40\%, the results are particularly bad for original \ac{s}, which is mainly overload by the biggest flows (elephant and rabbit), obtaining almost a null throughput. 
Finally, it is important to highlight that the enhanced \ac{s} shows smaller standard deviations than \ac{ovs}, although the values of \ac{ovs} are not bad either.

The main conclusion of this second test is the enhancements provided by BEBA make \ac{s} a feasible option for experiments dependent on higher performance. Indeed, the results of the \ac{s} switch are comparable to \ac{ovs}, reinforcing it as a reasonable option when modifications in the switch are required, or even when some features of OpenFlow are needed and not available in \ac{ovs}.

\section{Conclusions and Future Work}
\label{sec:future}

During the article, we have provided a guided overview of \ac{s}, trying to portray the importance of this software switch in \ac{sdn} environments, which are pivotal towards next-generation communication networks. 
We first introduced the history of the switch and presented its architectural design. Secondly, we described a set of selected use cases that leverage \ac{s} for diverse reasons: from easy customization to features missing in \ac{ovs}. The purpose was to highlight that, although \ac{ovs} may be thought as the king of software switches, \ac{s} can also be a good candidate for specific scenarios where \ac{ovs} is too complex (or almost impossible) to play with. 
Afterwards, we complemented the selected use cases with a comprehensive survey of works that also use \ac{s}, remarkable when the switch did not even had an official name and publication. 
Finally, we carried out an evaluation of \ac{s} vs. \ac{ovs} to prove that our switch has also a reasonable performance, greatly improved since the release of the original project. Researchers looking for a customized switch should carefully analyze the tradeoff between complexity and performance in \ac{ovs} and \ac{s}.

As future lines of work, we envision the growth of the community around \ac{s} and newer contributions for the switch. For this purpose, we have created a set of comprehensive guides, listed in Appendix~\ref{app:extguide}, to solve and help the work for researchers interested in the switch. 
Regarding the evolution of SBI protocols, the specifications of OpenFlow is currently stuck and the \ac{onf} is focusing now on the advanced programmability provided by the P4 language~\cite{Bosshart14} and P4 Runtime. Therefore, \ac{s} could join its efforts towards the adoption of this new protocol. 
In any case, we welcome any questions, suggestions or ideas to keep the \ac{s} community alive, and to do so, you can directly contact the team at the GitHub repository stated in~\cite{ofsoftswitch13}. 

\section*{Acknowledgements}
This work was partially supported by Ericsson Innovation Center in Brazil. Additional support is provided by CNPq (Conselho Nacional de Desenvolvimento Cient\'{i}fico e Tecnol\'{o}gico) grant numbers 310317/2013-4 and 310930/2016-2, by grants from Comunidad de Madrid through project TAPIR-CM (S2018/TCS-4496), and by the University of Alcala through project CCGP2017-EXP/001 and the ``Formaci\'{o}n del Profesorado Universitario (FPU)'' program.

\bibliographystyle{IEEEtran} %{spphys}%{splncs04}
\bibliography{paper}
\clearpage
\appendix
\section{Resources for Researchers and Developers}
\label{app:extguide}

\begin{itemize}

\item \textbf{Overview of the Switch's Architecture}. \\
\url{https://github.com/CPqD/ofsoftswitch13/wiki/Overview-of-the-Switch's-Architecture}

\item \textbf{Implementation Details}. \\
\url{https://github.com/CPqD/ofsoftswitch13/wiki/OpenFlow-1.3-Implementation-Details}

\item \textbf{How to Add a New OpenFlow Message}. \\
\url{https://github.com/CPqD/ofsoftswitch13/wiki/Adding-New-OpenFlow-Messages}

\item \textbf{How to Add a New Matching Field} \\
\url{https://github.com/CPqD/ofsoftswitch13/wiki/Adding-a-New-Match-Field}

\item \textbf{Frequently Asked Questions} \\
\url{https://github.com/CPqD/ofsoftswitch13/wiki/Frequently-Asked-Questions}
\end{itemize}

\end{document}